# GeoPipe: a Geo-distributed LLM Training Framework with enhanced Pipeline Parallelism in a Lossless RDMA-enabled Datacenter Optical Transport Network

Jun Dai
*State Key Lab of Information Photonics and Optical Communication*
*Beijing University of Posts and Telecommunication*
Beijing, P.R. China
winddomain@bupt.edu.cn

Xiaorun Wang
*State Key Lab of Information Photonics and Optical Communication*
*Beijing University of Posts and Telecommunication*
Beijing, P.R. China
wangxiaorun@bupt.edu.cn

Kexiong Fang
*State Key Lab of Information Photonics and Optical Communication*
*Beijing University of Posts and Telecommunication*
Beijing, P.R. China
2024140309@bupt.cn

Zheng Yang
*State Key Lab of Information Photonics and Optical Communication*
*Beijing University of Posts and Telecommunication*
Beijing, P.R. China
yangzheng2020@bupt.edu.cn

Yuefeng Ji*
*State Key Lab of Information Photonics and Optical Communication*
*Beijing University of Posts and Telecommunication*
Beijing, P.R. China
*corresponding email:jyf@bupt.edu.cn

Jiawei Zhang*
*State Key Lab of Information Photonics and Optical Communication*
*Beijing University of Posts and Telecommunication*
Beijing, P.R. China
*corresponding email:zjw@bupt.edu.cn

***Abstract*** **—The proliferation of Large Language Models (LLMs) with exponentially growing parameters is making cross-data center (DC) training an inevitable trend. However, viable strategies for extending single-DC training frameworks to multi-DC environments remain underdeveloped. We experimentally demonstrate, for the first time, a high-performance geo-distributed LLMs training framework across multiple DCs interconnected by a lossless, remote direct memory access (RDMA) enabled Datacenter Optical Transport Network (DC-OTN). An enhanced pipeline parallelism scheme is implemented within the Huawei's Ascend full-stack environment, which effectively eliminates the impact of cross-DC communication overhead on training efficiency. The overlapped computation and cross-DC communication is achieved with constraint cross-DC bandwidth and High Bandwidth Memory (HBM), reducing computation bubble ratio by up to ~78.91%.**

## I. Introduction

The training paradigm for large language models (LLMs) is increasingly shifting from single-datacenter deployments to geo-distributed multi-datacenter collaborations. This transition is motivated by three key factors: **(1) Energy supply**: The growing computational demands of LLMs are straining datacenter (DC) power infrastructures. Scaling a single DC is constrained by local power capacity and increased vulnerability to outages. Distributing workloads across multiple DCs helps alleviate energy pressure by leveraging distributed energy resources [1,2]. **(2) Data privacy**: Companies with limited computational re-sources rely on renting external Graphics Processing Units (GPU) to train proprietary models. To preserve data confidentiality, sensitive data must be shielded from public clouds. A viable approach is to retain data-dependent segments of the model, such as the input and output layers of a transformer-based LLM, within on-premises infrastructure, while offloading remaining computations to the public clouds. **(3) GPU fragmentation**: GPU vendors commonly operate geo-distributed DC regions and offer virtualized GPU leasing. When large-scale GPU allocations are required, it is often infeasible to concentrate all resources within a single DC region [3]. Instead, GPUs are provisioned across multiple regions, necessitating cross-DC coordination. Fig. 1 illustrates a geo-distributed LLM training scenario with the above considerations.

Training within single DC benefits from dedicated, high-bandwidth and flat intra-DC networks, where communication overhead is generally negligible. In contrast, geo-distributed training across multiple DCs typically relies on shared metro/core interconnects with limited bandwidth, resulting in communication stragglers, where slow links with high latency that substantially degrade training performance. Although some studies [4,5] proposed building dedicated, high-bandwidth and non-convergence cross-DC networks for distributed training, such solutions remain impractical due to network cost and inefficient bandwidth utilization [6]. Additionally, cross-DC networks exhibit lower resilience than intra-DC networks, experiencing increased packet loss due to the long-haul transmissions (ranging from tens to hundreds of kilometers) and physical links uncertainty. These conditions violate the lossless communication requirement essential for high-throughput remote direct memory access (RDMA) protocols.

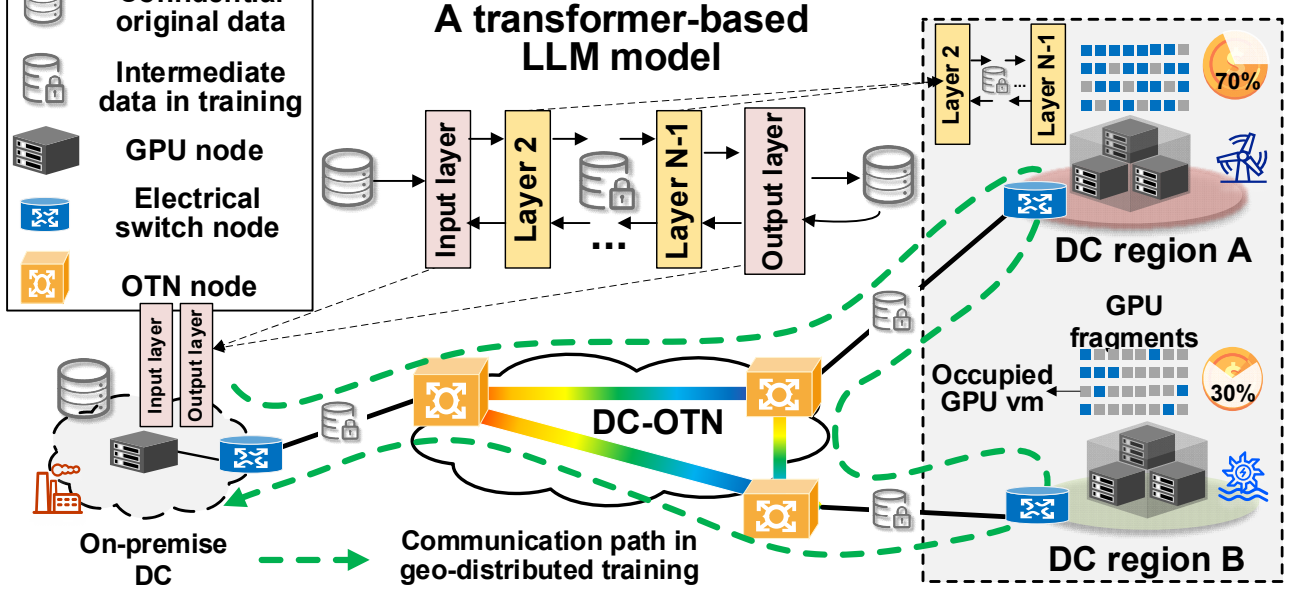

Fig. 1.A geo-distributed LLM training scenario.

In this paper, we propose GeoPipe, a comprehensive framework for high-performance geo-distributed LLM training. Our work makes the following key contributions: First, we design and implement an enhanced pipeline parallelism scheme that effectively overlaps computation with cross-DC communication to mitigate latency impact. Second, we deploy and integrate a datacenter optical transport network (DC-OTN) that provides deterministic latency and zero packet loss, enabling lossless long-haul RDMA transmission. Finally, we build a full-stack testbed comprising 64 Ascend Neural Processing Units (NPUs, same functions as GPUs for training) geo-distributed across three DCs, managed by Huawei's open-source MindSpeed framework [7]. The DCs are interconnected via DC-OTN nodes, with cross-DC links operating at a total line rate of 3.2 Tbps over 120 km. Evaluated on the Llama-2 13B and 72B models under constrained cross-DC bandwidth and HBM capacity, GeoPipe achieves a 78.91% reduction in the computation bubble ratio compared to the conventional 1F1B scheme.

## II. Problem Statement and Analysis

### A. Communication Characteristics of LLM Training

Large-scale training of LLMs employs advanced parallelization strategies to distribute computational workloads across GPU clusters. These strategies primarily include Data Parallelism (DP), Tensor Parallelism (TP), Pipeline Parallelism (PP), and Expert Parallelism (EP), which are frequently combined into a Hybrid Parallelism (HP) framework [8]. A representative iteration of a three-dimensional (3D) hybrid parallel scheme integrating TP, PP, and DP is illustrated in Fig. 2(a). Each complete iteration encompasses forward computation, backward computation, and interleaved communication phases for TP, PP, and DP. This workflow corresponds to the data trajectory shown in Fig. 1, where activations propagate forward from the first layer to the final layer followed by gradient flow in the reverse direction.

Under the DP strategy, independent model replicas are maintained and trained on different datasets, with each DP group utilizing a dedicated set of GPUs. Weight synchronization across replicas occurs only at the end of each iteration through an all-reduce operation (denoted by ③ in Fig. 2(a)). All other computational and communication steps remain independent and structurally similar across DP groups.

PP partitions the model layer-wise into stages, typically with multiple layers composing one stage (equivalent to one GPU worker). The dataset (i.e., the input of a general matrix multiply) for each iteration is divided into multiple micro-batches. After completing all forward (or backward) computations for a micro-batch within a stage, cross-GPU PP communication is required (denoted by ② in Fig. 2(a)), with each stage processing only one micro-batch at any given time. Label ② demonstrates the PP communication process for micro-batch 1 in a classic 1F1B PP scheme (with *m* micro-batches per iteration). After stage0 completes its forward computation, it transfers activations to the downstream stage1 via **send-recv** communication, and this process continues sequentially until gradients eventually flow back to stage0.

In contrast, TP decomposes individual layers of the model into smaller sub-operations distributed across multiple GPUs. Each layer-wise computation is followed by a synchronization step—implemented via all-reduce (as indicated by label ① in Fig. 2(a)). Consequently, every micro-batch requires frequent TP synchronization within each stage, resulting in high communication frequency and volume.

In summary, among these parallelization strategies, TP involves frequent, fine-grained, and high-volume communication, making it particularly sensitive to latency and bandwidth constraints. These characteristics render it unsuitable for cross-DC deployment. PP, on the other hand, employs a point-to-point communication pattern with predictable volume and significantly lower frequency, making it better suited for geo-distributed training environments. Although DP could in principle be deployed across DCs [9], it is not considered in this work.

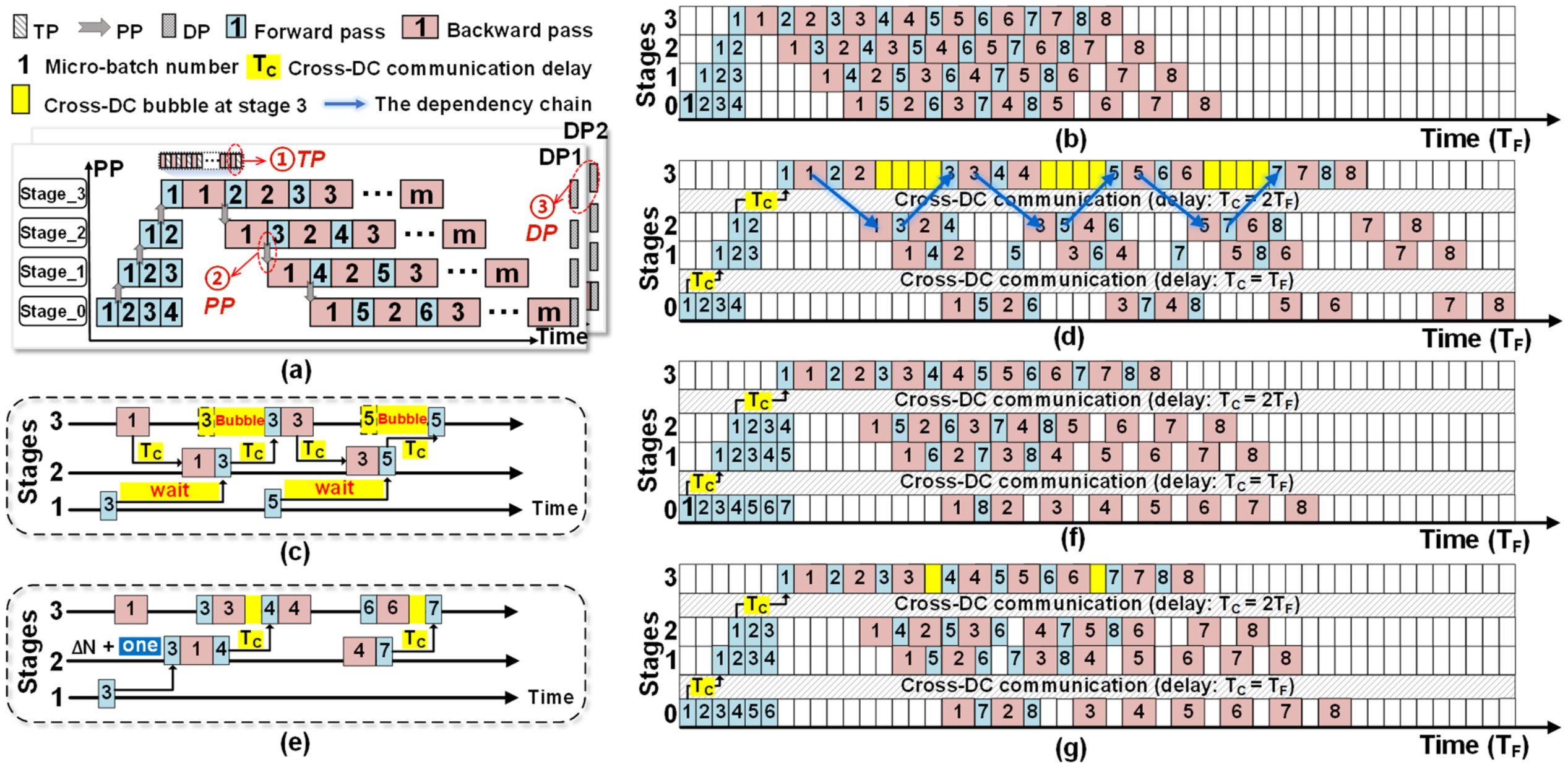

Fig. 2.(a) Workflow of a 3D hybrid parallelism (TP, PP, DP) training iteration; (b) Examples of pipeline parallelism runtime with 4 stages and its first 8 micro-batches: inter-DC PP communication with ideal zero bubbles; (c) Impact of cross-DC communication latency on pipeline; (d) Cross-DC PP communication with introduced cross-DC bubbles; (e) Optimized schedule for cross-DC bubble absorption; (f) Optimized cross-DC PP communication without HBM constraint; (g) Optimized cross-DC PP communication with HBM constraint.

### B. Problem Statement: The Cross-DC Bubble for geo-distributed training

Fig. 2(b) illustrates an example of a classic micro-batch-based 1F1B PP scheme runtime with 4 stages and its first 8 micro-batches without considering cross-DC communication delays, where forward/backward computation and TP communication are abstracted into unified pass durations (denoted as $\boldsymbol{T_F}$ for forward and $\boldsymbol{T_B}$ for backward). To analyze geo-distributed training, we extend this scheme by placing stage3 in an on-premises DC while hosting stage1 and stage2 within a GPU provider's DC. This arrangement makes the stage2→stage3(stage3→stage2) link a cross-DC communication path, which introduces non-negligible latency ($Tc$) comprising both distance-related ($L$) and bandwidth-related ($BW$) components.

As shown in Fig. 2(c), the strict 1F1B rule during the steady phase creates a critical dependency. The forward pass of micro-batch 3 at stage2 must await the completion of the backward pass of micro-batch 1 at the same stage. However, this backward pass itself depends on the backward pass of micro-batch 1 from the preceding stage3 and the subsequent cross-DC communication. The latency $Tc$ in this dependency chain thus induces a computation bubble at stage3.This bubble does not remain isolated; it propagates downstream through the backward pass of micro-batch 3, eventually causing a stall in micro-batch 5 and triggering cascading pipeline stalls. We define this bubble, originating from cross-DC communication, as the "**cross-DC bubble**". Critically, this bubble directly degrades training efficiency by introducing idle time into the computational pipeline, which accumulates and significantly increases the average iteration time.

The total cross-DC bubble time at the final stage is determined by the product of the bubble duration per occurrence and its frequency. This leads to the following theoretical formulation:

$$T_{cross-DC\ bubble} = g(\overrightarrow{T_C}) \times f(m) \tag{1}$$

here the function $g(\overrightarrow{T_C})$ represents the bubble duration per occurrence, which depends on the cross-DC latency vector $\overrightarrow{T_C}$ between adjacent stages and their geo-distribution. The term $f(m)$ denotes the frequency of this bubble, corresponding to the total number of micro-batches $m$ per training iteration.

Based on this model, we extend the scenario in Fig. 2(b) to a three-DC interconnected setup, as depicted in Fig. 2(d). In this configuration, a cross-DC communication delay of one TF unit exists between stage0 and stage1, while a delay of two TF units is introduced between stage2 and stage3. Our analysis reveals that the cross-DC bubble is primarily attributable to the longer latency between stage2 and stage3. We identify this bottleneck as the critical path, which we term the dependency chain—the inter-stage link with the maximum latency that governs the overall schedule. This dependency chain is illustrated by the blue arrows in Fig. 2(d).

### C. Communication-Computation Overlap Model

To mitigate this, we increase the scheduling lead of forward passes between stage $i$ and stage $i$+1 along the dependency chain during the warm-up phase. This approach helps absorb more cross-DC bubbles caused by communication delays $Tc$. As shown in Fig. 2(e), advancing micro-batch 3 in stage2 eliminates its dependency on the delayed backward pass of micro-batch 1, thereby allowing micro-batch 4 to occupy the original time slot of micro-batch 3. As a result, a cross-DC bubble of size $2Tc$ is absorbed by the combined duration of the forward ($T_F$) and backward ($T_B$) passes of micro-batch 3.

Based on this insight, the impact of cross-DC communication delay can be masked by overlapping it with computation. The effectiveness of this masking correlates with the increase in forward passes during the warm-up phase (denoted as $\Delta N$) and the execution times of forward and backward passes ($T_F$ and $T_B$, respectively). We define $T_{overlap}$ as the amount of cross-DC communication delay that is successfully hidden within computation, and $Tc'$ as the effective latency between stage $i$ and stage $i$+1 after masking. This relationship is captured as follows:

$$T_C{}' = T_C - T_{overlap} \tag{2}$$

$$T_{overlap} = \Delta N \times \frac{(T_F+T_B)}{2} \tag{3}$$

Note that $T_F$ and $T_B$ vary with sequence length, micro-batch size, and GPU throughput—all measured in both simulation and experimental environments. As shown in Fig. 4(b) (down), when the total number of tokens per micro-batch (i.e., the product of sequence length and micro-batch size) is small, computational throughput does not peak and instead introduces additional idle time, causing $T_F$ and $T_B$ to decrease anomalously as the token count per micro-batch increases. Beyond a certain token volume, computational throughput stabilizes near its peak, and both $T_F$ and $T_B$ grow nearly linearly with further increases in token count. Meanwhile, $\Delta N$ is constrained by HBM bound and token count, as illustrated in Fig. 4(b) (top). The HBM usage increases with the token count per micro-batch. For a fixed token count, HBM usage scales linearly with $\Delta N$.

As shown in Fig. 3(f), our algorithm achieves the complete elimination of cross-DC bubbles caused by latency when operating without HBM constraints. Under HBM constraints, the optimization outcome in Fig. 3(g) demonstrates that our method remains highly effective. Within HBM limitations, the system iteratively refines the dependency chain; for example, after optimizing the Stages 2-3 link to reduce effective latency ($T_{overlap}$), the critical path shifts, making the Stages 0-1 link the new dependency chain.

Our communication-computation overlap model is predicated on the deterministic latency and zero packet loss guaranteed by the DC-OTN. This allows us to accurately model the cross-DC latency $T_C$ not as a stochastic variable, but as a predictable parameter, which is fundamental to the reliability of our scheduling algorithm. The formulation of $T_{overlap}$ and the subsequent reduction in effective latency $T_C'$ are only possible under such a stable optical transport layer. Conversely, increasing $T_{overlap}$ allows for a larger communication window for cross-DC data transfer. For a fixed transmission distance $L$, this expanded window reduces the demand on bandwidth $BW$, thereby lowering communication costs and improving the utilization efficiency of the OTN.

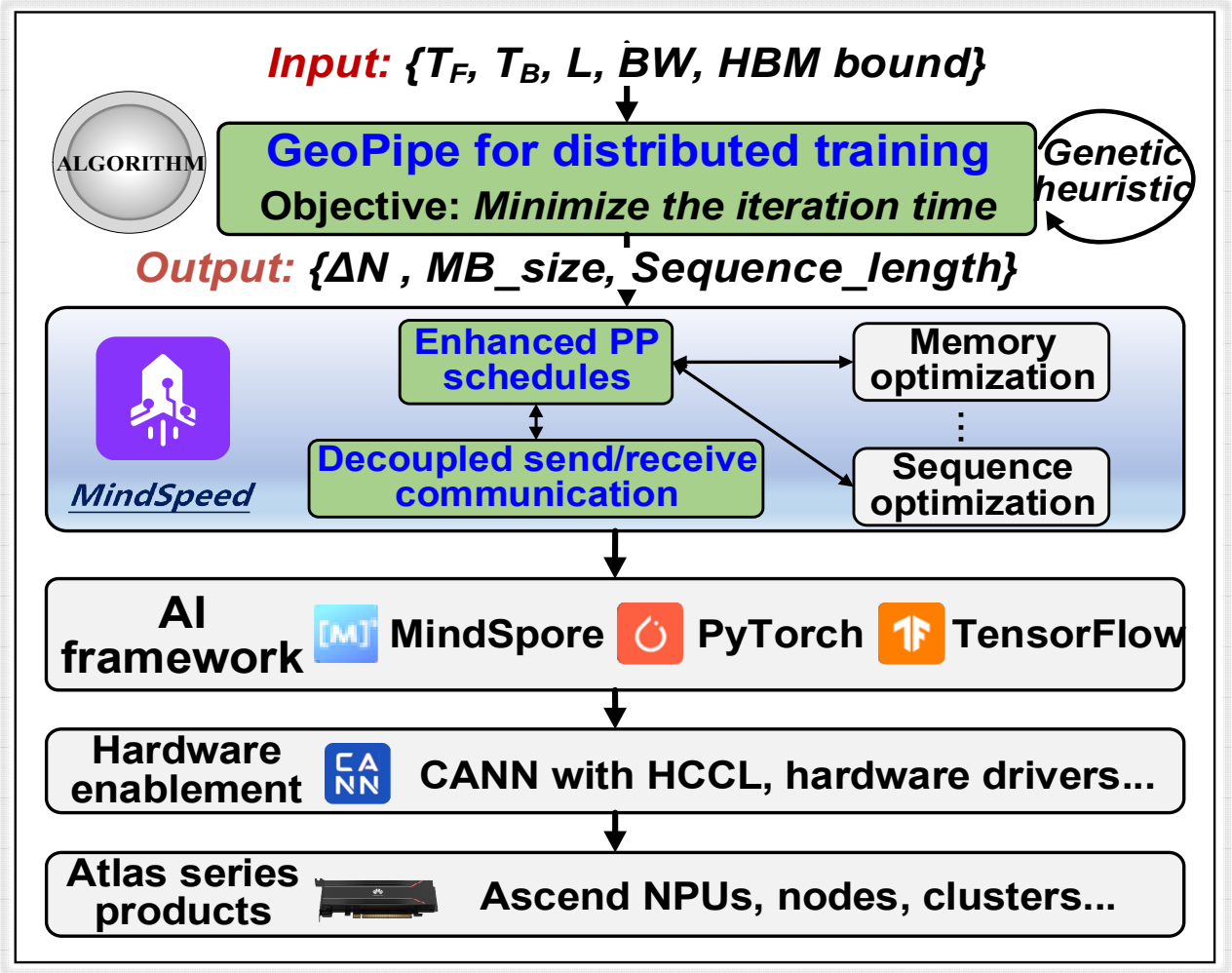

Fig. 3.GeoPipe framework based on Ascend full-stack AI environment.

This method is complementary to advanced pipelines like ZeroBubble [10], collectively working to minimize the overall cross-DC bubble time.

## III. GeoPipe-Enabled Geo-Distributed LLM Training Framework

GeoPipe is designed to identify an optimal PP configuration to mitigate the impact of cross-DC bubbles. The algorithm takes computation time, HBM bound, cross-DC latency and bandwidth as inputs, and iteratively adjusts micro-batch size, sequence length and the number of forward passes ($\Delta N$) through genetic heuristic. This optimization continues until the shortest iteration time for the same number of total tokens is achieved. Fig. 3 shows the GeoPipe-enabled distributed training based on Ascend full-stack AI environment [11].

MindSpeed-LLM is a high-performance, full-stack LLM training framework developed for supporting Huawei's Ascend AI accelerator (like Ascend 910A/B/C series). It integrates advanced algorithms that can automatically split a large model across multiple NPUs using various parallel strategies, such as DP, TP, PP, and their various combinations. MindSpeed-LLM implements advanced memory management and communication scheduling optimizations, which provide the necessary infrastructure for effective GeoPipe deployment.

The GeoPipe engine is implemented in Python and running on top of MindSpeed-LLM framework to support geo-distributed LLM learning. GeoPipe is a pipeline scheduler that generates adaptive PP configurations for geo-distributed training scenarios, and collaborates with a hybrid-parallel orchestrator in Mind-Speed-LLM to determine the communication topology and constructs TP, DP, and PP communication groups. To meet the GeoPipe's requirement, we decouple the send/receive operation in PP as a non-blocking communication operation, where send operation do not need to wait its corresponding receive end to execute the next operation (e.g., computational operation). This modification can maximize the communication overlap.

## IV. Experimental Setup and Data Analysis

Our testbed comprises eight compute nodes, each equipped with eight Ascend 910B NPUs featuring 64 GB of HBM. Intra-node communication employs a full-mesh CXL bus providing 200 Gbps per NPU pair, while inter-node communication is facilitated by RDMA NICs delivering 200 Gbps per NPU port. The 64 NPUs are distributed across three geo-distributed DC regions. Within each region, compute nodes are interconnected by the two-layered electrical switches operating at 400 Gbps per port. The regions are connected via a DC-OTN backbone. Each DC-OTN device incorporates a large buffer and embedded congestion control mechanisms like Priority Flow Control (PFC), ensuring lossless RDMA transmission. Cross-DC bandwidth can be dynamically adjusted through the control of the spine switch port rates. This process is monitored in real-time by Tesgine, a devise deployed by Huawei for data stream monitoring. Fig. 4(a) details the experimental topology and configurations. We evaluate GeoPipe on the Llama-2 13B and 72B models using a hybrid parallelism strategy across 64 NPUs in three DC regions. Specifically, we employ a TP degree of 8, confined to intra-node communication, and a PP degree of 8, where only PP traffic traverses the cross-DC network. GeoPipe is benchmarked against the classical 1F1B scheduling strategy.

Validation of Lossless RDMA-enabled OTN Transmission. As the foundation of our geo-distributed system, the lossless characteristic of the DC-OTN is critical. To rigorously stress-test this, we deliberately introduced congestion by reducing a spine switch port rate from 400 Gbps to 200 Gbps at time $t_a$. We then monitored the OTN throughput in real-time using Tesgine and tracked packet retransmissions via the HCCN_tool. As shown in Fig. 4(c), with PFC enabled, the OTN throughput remains stable and no packets are retransmitted, validating the deterministic, lossless transport that underpins our long-haul RDMA. This result directly confirms that our infrastructure meets the prerequisite stated in the introduction for high-throughput RDMA protocols, thereby enabling a stable foundation for GeoPipe's scheduling optimizations.

High-performance Training Efficiency. Fig. 4(d) evaluates the core benefit of GeoPipe—its ability to minimize the "cross-DC bubble" defined in our problem statement. Under varying cross-DC bandwidth and model parameters, GeoPipe demonstrates substantial performance gains. Specifically, it achieves a maximum reduction of 78.91% in the computation bubble ratio and a corresponding ~9.1% improvement in average iteration time compared to the classic 1F1B scheduler. This empirically validates the effectiveness of our communication-computation overlap model. Furthermore, a key finding emerges: for any given model size and transmission distance, there exists a "**cross-DC bandwidth optimization point**". This point, which balances bandwidth cost with training efficiency, aligns closely with the peak of the bubble ratio curve. This discovery provides a crucial, data-driven guideline for resource provisioning in geo-distributed training clusters, moving beyond mere heuristic configuration.

System-Wide Efficiency. The output from the MindStudio profiler in CANN, shown in Fig. 3(e), offers a microscopic view of GeoPipe's scheduler in action. The timeline clearly shows dense overlap between computation and cross-DC communication, visually confirming that our enhanced pipeline parallelism and non-blocking communication implementation successfully mask latency. This translates the theoretical $T_{overlap}$ from our mathematical model into observed system behavior, completing the narrative from algorithm design to practical realization.

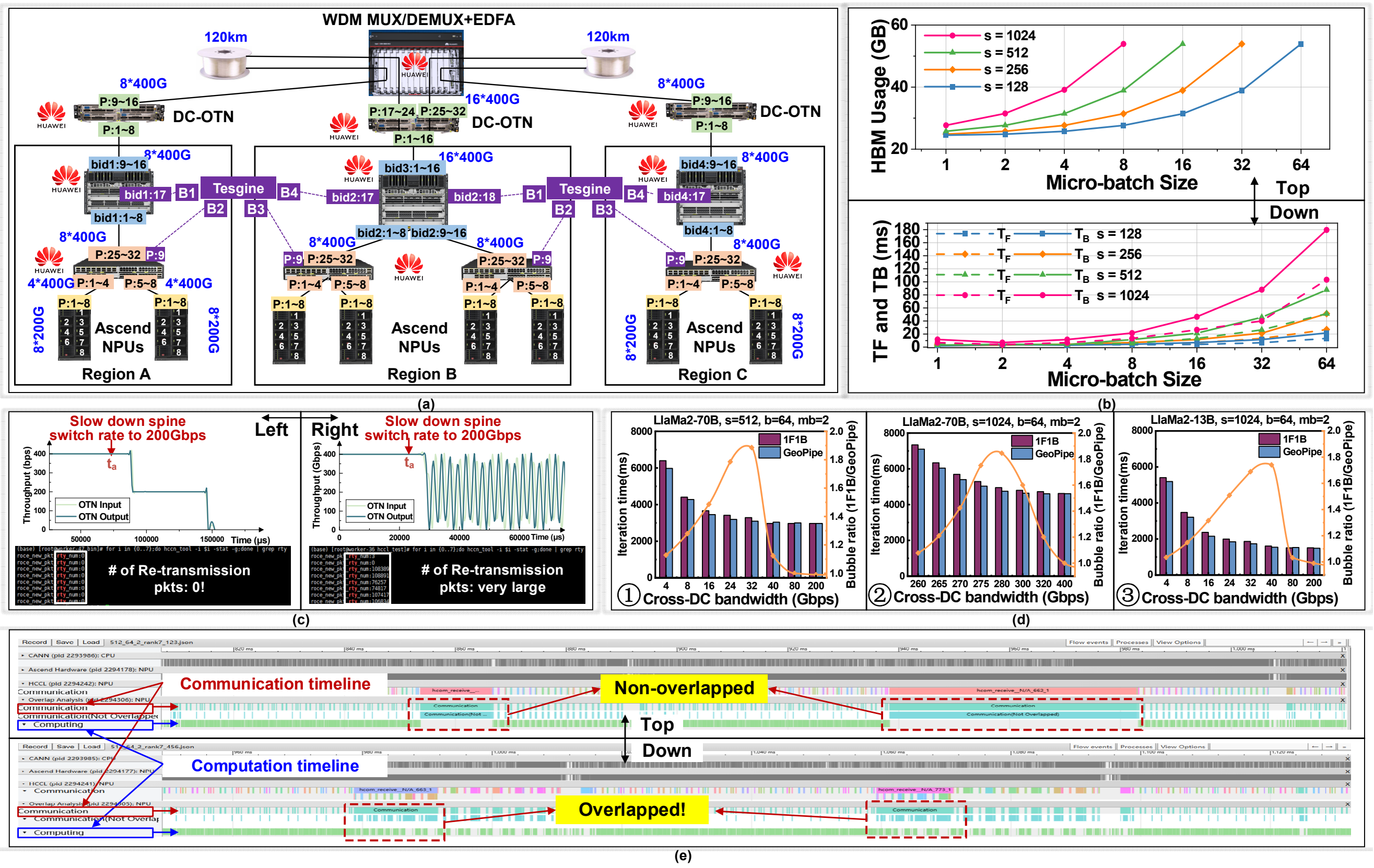

Fig. 4.(a) Experimental topology and configurations; (b) $T_F$, $T_B$ and HBM usage under different micro-batch size and sequence length; (c) OTN throughput and packet retransmission w/ (left) and w/o (right) PFC; (d) Geo-Pipe vs. classic 1F1B under varying cross-DC bandwidth and model parameters; (e) Overlap between computation and cross-DC communication for 1F1B (top) and GeoPipe (down).

## V. Conclusion

We have presented GeoPipe, a framework that co-designs distributed LLM training with a deterministic DC-OTN. Our work demonstrates that the performance predictability of optical networks is not merely an enhancement but a prerequisite for efficient geo-distributed LLM training. The 78.91% reduction in computation bubble ratio achieved on our testbed signals the dawn of optical network-aware computing. Future work will explore tighter integration between the training scheduler and the optical control plane for dynamic resource orchestration.

## Acknowledgment

This work was supported by the National Key R&D Program of China (No. 2022YFB2903700 and 2024YFB2908303), National Natural Science Foundation of China (62271078), and the Fund of State Key Laboratory of IPOC (No. IPOC2025ZZ03). We would like to extend our sincere gratitude to the colleagues at Huawei for their support and suggestions throughout the development of the experimental platform.